\documentclass[aps,pra,twocolumn,amsmath,amssymb,showpacs,superscriptaddress]{revtex4}

\usepackage{graphicx}
\usepackage{bm}

\begin{document}

\title{Possibility of a cosmological gravimetry in general relativity}

\author{V. I. Yudin}
\email{viyudin@mail.ru}
\affiliation{Novosibirsk State University, ul. Pirogova 2, Novosibirsk, 630090, Russia}
\affiliation{Institute of Laser Physics SB RAS, pr. Akademika Lavrent'eva 13/3, Novosibirsk, 630090, Russia}
\affiliation{Novosibirsk State Technical University, pr. Karla Marksa 20, Novosibirsk, 630073, Russia}
\author{A. V. Taichenachev}
\affiliation{Novosibirsk State University, ul. Pirogova 2,
Novosibirsk, 630090, Russia}
\affiliation{Institute of Laser Physics SB RAS, pr. Akademika
Lavrent'eva 13/3, Novosibirsk, 630090, Russia}

\date{\today}

\begin{abstract}
In the framework of the parametrized post-Newtonian (PPN) formalism, we substantiate an idea according to which we can measure the value of the cosmological gravitational potential $\Phi$ at the location of the Solar System, which is formed by all the matter of the Universe (including dark matter).
This paradoxical result is based on the fact that the PPN formulation of general relativity is not invariant with respect to the transformation $U({\bf r})\,\rightarrow\,U({\bf r})+C$, where $U({\bf r})$ is a gravitation potential and $C$ is an arbitrary real constant (this is due to the nonlinearity of Einstein equations, first of all).
 Starting from the cosmological description of the Universe within the framework of the standard general relativity (i.e., using initial PPN parameters $\gamma_0=\beta_0=1$ in the cosmological reference frame), we show that from the viewpoint of the local (laboratory) observer there is a renormalization of the PPN parameters $\gamma$ and $\beta$: $|\gamma-1|\sim |\Phi|/c^2$ and $|\beta-1|\sim |\Phi|/c^2$. This leads to the estimates: $|\beta-1|>10^{-6}$ and $|\gamma-1|>10^{-6}$ for gravitational experiments in the Solar System. In addition, we have found an unique post-Newtonian model, which is invariant with respect to the transformation $U({\bf r})\,\rightarrow\,U({\bf r})+C$ and therefore is insensitive to the cosmological potential $\Phi$. The most unusual is that for this model we always have $\gamma=\beta=1$ from the viewpoints of the cosmological and any laboratory observers.
\end{abstract}


\maketitle
\section{Introduction}

More than a hundred years have passed since the advent of general relativity (GR), developed by A. Einstein at the beginning of the 20th century \cite{Einstein_1907,Einstein_1911,Einstein_1915,Einstein_1916}. However, all this time GR continues to be an area of active research, both theoretical and experimental (see the reviews \cite{Will_2014,Tur_2008}). The verification of general relativity was carried out with increasing accuracy, and today it successfully explains almost all the available data from numerous experiments. The revolution in the experimental testing of GR has occurred over the past 50 years due to great progress in various fields of science and technology. First of all, this is due to the development of space research, the emergence of high-precision space navigation methods for spacecrafts, a significant improvement in the accuracy of astronomical observations, the development of laser range-finding of the Moon, etc.

Recent advances in applied physics have made available new instrumentation and technologies. For example, a new generation of quantum sensors (ultra-stable atomic clocks, accelerometers, gyroscopes, gravimeters, gravity gradiometers) significantly increases the potential of the modern measurement base, allowing us to achieve extremely high levels of accuracy in testing the fundamentals of modern physics, in general, and GR, in particular. The latest achievements in this area include the results of Refs.~\cite{Delva_2018,Herrmann_2018}, in which authors report on successful tests (at the level of $10^{-5}$) of the gravitational redshift and thus of local position invariance, as an integral part of the Einstein equivalence principle, which is the foundation of GR and all metric theories of gravitation.

Another main direction in the verification of general relativity is the experimental measurement of the so-called post-Newtonian parameters. This method is based on the phenomenological parametrization of the metric tensor of the gravitational field and is called the parametrized post-Newtonian (PPN) formalism, which is suitable for working with a wide class of metric theories, including GR as a special case. Within the PPN formalism, the specific metric theory of gravity is completely characterized by ten PPN parameters \cite{Will_2014,Will_2018,Moyer1,Moyer2}. Formalism uniquely predicts the values of these parameters for each particular theory. Gravity experiments can be also analyzed in the framework of the PPN formalism. In this case, the experimental results make it possible to determine the values of the post-Newtonian parameters and, therefore, to conclude in favor of a particular theory.

Of particular importance are two PPN parameters ($\beta$,$\gamma$), which have a clear physical meaning: $\gamma$ is a measure of the curvature of space-time created by an unit rest mass; $\beta$ is a measure of the nonlinearity of the law of superposition of gravitational fields. For example, the GR in the standard PPN calibration yields $\beta=\gamma=1$, and all other eight parameters disappear.

To date, much empirical data have already been accumulated on the measurement of post-Newtonian parameters ($\beta$,$\gamma$), obtained both using ground-based high-precision astrometric equipment and in experiments using spacecraft (see reviews \cite{Will_2014,Tur_2008}). Such experiments include: measuring the anomalous perihelion shift of Mercury's orbit, measuring the deflection of light near the Sun, the time delay of radar signals sent across the Solar System past the Sun to a planet or satellite and returned to the Earth (due to the gravitation of the Sun), laser ranging of the Moon, and some others. Analyzing the obtained results, we can with high probability assert that within the Solar System there is: $|\beta-1|<10^{-4}$ and $|\gamma-1|<10^{-4}$. On the one hand, this indicates the high reliability of GR predictions. On the other hand, the achieved level of accuracy is still not so high as to guarantee, for example, the incorrectness of some alternative gravity theories for which $\beta\neq 1$ and $\gamma\neq 1$ (see reviews \cite{Will_2014,Tur_2008}).

In this paper, we study the influence of the cosmological background on the value of post-Newtonian parameters ($\beta$,$\gamma$) in the interpretation of gravitational experiments within the Solar system. By cosmological background, we mean the gravitational contribution from all the celestial bodies of the Universe, located far beyond the Solar System (including dark matter and dark energy). Our theoretical scheme is following. In the cosmological reference frame, we start directly with a nonlinear solution of the Einstein equations including post-Newtonian and post-post-Newtonian terms. Then, we show how this solution will look in the reference frame of the Laboratory Observer near small gravitating bodies.

Using the general PPN formalism, we find that for a local observer (i.e., for an experimentalist in the Solar System) the values of post-Newtonian parameters ($\beta$,$\gamma$) differ from their initial values ($\beta_0$,$\gamma_0$), which appear in the theoretical description of gravitation from the viewpoint of the virtual Cosmological Observer. Moreover, information on the cosmological background is contained in the quantities $(\beta-\beta_0)$ and $(\gamma-\gamma_0)$. In particular, in the case of GR ($\beta_0=\gamma_0=1$) we get a lower bound: $ |\beta-1|>10^{-6}$ and $|\gamma-1|>10^{-6}$ in the Solar System. We also find a previously unknown post-Newtonian model of gravitation, which is insensitive to the cosmological background (in this model, we have $\gamma=\gamma_0=\beta=\beta_0=1$ for any Laboratory Observer). The obtained results can be considered a theoretical basis for experimental cosmological gravimetry.

\section{General Relativity Formalism}
Let us consider some basic formulas of general relativity, which are necessary for our further considerations. In this case, the physical picture of gravitation is described by a metric in four-dimensional space-time, when the relationship between two infinitely close space-time events is given using the interval:
\begin{equation}\label{int}
ds^2=g^{}_{jk}(\vec{x})dx^j dx^k\quad (j,k=0,1,2,3)\,,
\end{equation}
where $g^{}_{jk}(\vec{x})=g^{}_{jk}(t,{\bf r})$ is the covariant metric tensor depending on the coordinates of the four-vector $\vec{x}=\{ x^0=ct,x^1,x^2,x^3\}=\{ct,{\bf r}\}$. This metric tensor defines the standard equations of the geodesic line along which the test body of small mass moves:
\begin{equation}\label{geod_s}
\frac{\partial^2 x^j}{\partial s^2}+\Gamma^{j}_{mn}\frac{\partial x^m}{\partial s}\frac{\partial x^n}{\partial s}=0,\quad (j,m,n =0,1,2,3)\,,
\end{equation}
where $\Gamma^{k}_{ij}$ are the Christoffel symbols:
\begin{equation}\label{Krist}
\Gamma^{j}_{mn}=\frac{1}{2}g^{jk}\left(\frac{\partial g^{}_{km}}{\partial x^n}+\frac{\partial g^{}_{kn}}{\partial x^m}-\frac{\partial g^{}_{mn}}{\partial x^{k}}\right), \;\; (g^{jl}g^{}_{lk}=\delta^{j}_{k})\,.
\end{equation}
The equations (\ref{geod_s}) related to the coordinate time $t$ can be rewritten in the following form (see, for example, Ref.~\cite{Brumberg_1991}):
\begin{equation}\label{geod_x0}
\frac{d^2 x^j}{dt^2}+\Gamma^{j}_{mn}\frac{d x^m}{dt}\frac{d x^n}{dt}-c^{-1}\Gamma^{0}_{mn}\frac{d x^m}{dt}\frac{d x^n}{dt}\frac{d x^j}{dt}=0\,.
\end{equation}
Note that there are only three independent equations for $j=1,2,3$, because Eq.~(\ref{geod_x0}) for $j=0$ corresponds to the identity: $0=0$. In the absence of gravitation, the tensor $g_ {jk}$ describes the metric of Minkowski space-time for the special relativity:
\begin{equation}\label{Mink}
g_{jk}^{(M)}=\left(
  \begin{array}{cccc}
    1 & 0 & 0 & 0 \\
    0 & -1 & 0 & 0 \\
    0 & 0 & -1 & 0 \\
    0 & 0 & 0 & -1 \\
  \end{array}
\right),
\end{equation}
in which geodesic lines correspond to a uniform rectilinear motion: ${d^2 {\bf r}}/{dt^2}=0$.

For simplicity, we will consider the case of a static Universe, i.e. when the distribution of masses in the Universe does not change over time. Moreover, the metric tensor $g^{}_{jk}$ will be described in the initial flat time space $\{t,{\bf r}\}_\text{Cosm}$, which we associate with the ``Cosmological Observer''. In this case, to describe a weak gravitational field, the following expression for the interval is a good approximation:
\begin{equation}\label{g_usual}
ds^2 =g^{}_{00}({\bf r})\,(c\,dt)^2-g^{}_{11}({\bf r})\,d{\bf r}^2\,,
\end{equation}
when there is a diagonal spatially isotropic metric tensor:
\begin{equation}\label{g_gen}
g_{jk}^{}=\left(
  \begin{array}{cccc}
   g^{}_{00}({\bf r}) & 0 & 0 & 0 \\
    0 & -g^{}_{11}({\bf r}) & 0 & 0 \\
    0 & 0 & -g^{}_{11}({\bf r}) & 0 \\
    0 & 0 & 0 & -g^{}_{11}({\bf r}) \\
  \end{array}
\right),
\end{equation}
and the corresponding dual tensor $g^{jk}$:
\begin{equation}\label{dual_g}
g^{jk}=\left(
  \begin{array}{cccc}
   \frac{1}{g^{}_{00}({\bf r})} & 0 & 0 & 0 \\
    0 & -\frac{1}{g^{}_{11}({\bf r})} & 0 & 0 \\
    0 & 0 & -\frac{1}{g^{}_{11}({\bf r})} & 0 \\
    0 & 0 & 0 & -\frac{1}{g^{}_{11}({\bf r})} \\
  \end{array}
\right).
\end{equation}
For such a metric tensor, the equation of the geodesic line (\ref{geod_x0}) in the nonrelativistic limit on the velocity of the test body ($|\dot{{\bf r}}|/c$$\ll$1) leads to the following main approximation:
\begin{equation}\label{geod_00}
\frac{d^2 x^j}{dt^2}\approx -\Gamma^{j}_{00}c^2=-\frac{c^2}{2g^{}_{11}({\bf r})}\frac{\partial g^{}_{00}({\bf r})}{\partial x^j}\,,\quad (j=1,2,3)\,,
\end{equation}
which can be written in the vector form:
\begin{equation}\label{geod_vec}
\frac{d^2 {\bf r}}{d t^2}=-\frac{c^2}{2g^{}_{11}({\bf r})}\nabla_{\bf r}\,{g^{}_{00}({\bf r})}\,,
\end{equation}
where $\nabla_{\bf r}$=${\bf e}_1(\partial /\partial x^1)$$+$${\bf e}_2(\partial /\partial x^2)$$+$${\bf e}_3(\partial /\partial x^3)$ is the standard spatial gradient operator (${\bf e}_j$ is a unit vector along $j$th Cartesian axis). Note that Eq.~(\ref{geod_vec}) will play a key role in our further considerations.

\section{Concept of Cosmological Gravitational Potential}

\begin{figure}[t]
\centerline{\scalebox{0.35}{\includegraphics{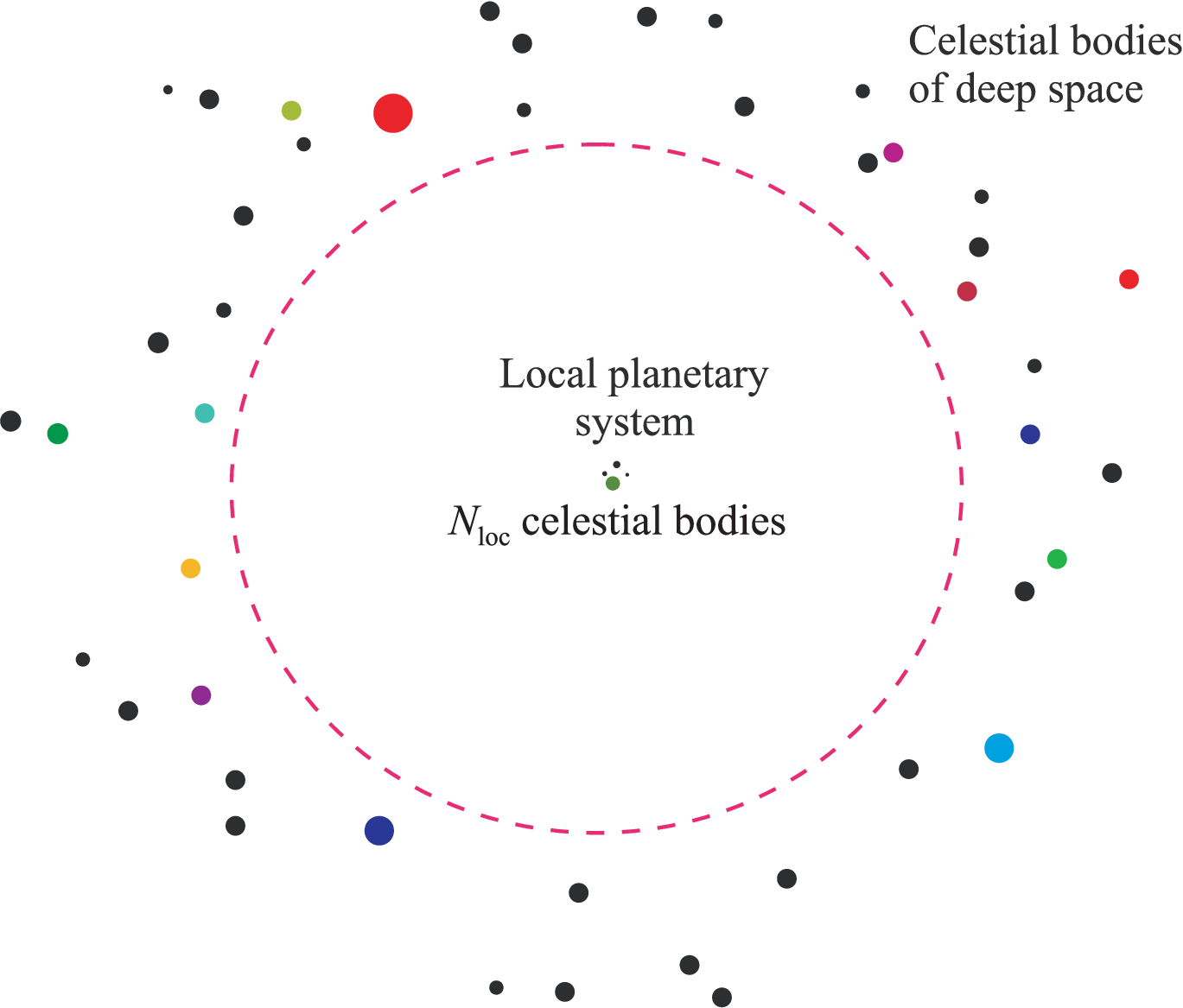}}}\caption{
A conditional illustration of the local planetary system ($N_\text{loc}$ bodies in the center) and other bodies of the Universe (on the outskirts). In our paper, we describe the gravitation inside the local planetary system.} \label{Fig1}
\end{figure}

In this paper, we assume that the source of gravitation in a static Universe is an arbitrary number of point masses $M_1,\,M_2,...,M_n,...$, which are stationary in space, i.e. we will neglect the movement of celestial bodies. In this case, the Newtonian gravitational potential is described in standard way:
\begin{equation}\label{Newton_pot}
U({\bf r})= -\sum_{n}\frac{\Gamma_0  M_n}{|{\bf r}-{\bf R}_n|}\,,
\end{equation}
where $\Gamma_0$ is the gravitational constant from the viewpoint of the Cosmological Observer, and ${\bf R}_n$ is the coordinate of a point source with mass $M_n$. Note that in our work we will follow the classical definition of the Newtonian attraction potential, which has a negative value, $U({\bf r})<0$.

We will consider a small area of the Universe far from very massive bodies when the condition of small gravitational potential, $|U({\bf r})/c^2|$$\ll$1, is fulfilled. As a specific example, we will explore the spatial region within the Solar system. In this case, the Newtonian potential (\ref{Newton_pot}) can be represented as follows:
\begin{equation}\label{U_structure}
U({\bf r})=U_\text{loc}({\bf r})+\Phi({\bf r})\,,
\end{equation}
where the local potential
\begin{eqnarray}\label{U_local}
U_\text{loc}({\bf r})&=&-\sum_{n=1}^{N_\text{loc}}\frac{\Gamma_0 M_n}{|{\bf r}-{\bf R}_n|} \\
&=&U_\text{E}({\bf r})+U_\text{S}({\bf r})+U_\text{M}({\bf r})+U_\text{J}({\bf r})+...\nonumber
\end{eqnarray}
includes summation over the indices $\{1,2,...,N_\text{loc}\}$, which number only the sources of gravitation inside the planetary system under consideration. In our particular case, $U_\text{loc}({\bf r})$ includes the Newtonian potentials of the Earth $U_\text{E}({\bf r})$, the Sun $U_\text{S}({\bf r})$, the Moon $U_\text{M}({\bf r})$, the Jupiter $U_\text{J}({\bf r})$ and other bodies of the Solar system. The second contribution in Eq.~({\ref{U_structure}}) corresponds to the cosmological potential $\Phi({\bf r})$:
\begin{equation}\label{CP}
\Phi({\bf r})=-\sum_{n>N_\text{loc}}\frac{\Gamma_0 M_n}{|{\bf r}-{\bf R}_n|}\,,
\end{equation}
including potentials from all other bodies in the Universe, which are at huge distances from the Solar System, as well as the effects of dark matter.

Let us now evaluate the order of magnitude of the various contributions. Near the Earth's surface, we have an estimate of the Earth's potential $|U_\text{E}|/c^2$$\approx$$\,0.7\times 10^{-9}$, and the estimate $|U_\text{S}|/c^2$$\approx$$\,10^{-8}$ corresponds to the potential of the Sun in the Earth's orbit. To estimate the lower bound of the cosmological potential $\Phi$, we consider the orbital motion of the Solar system around the center of the Galaxy, which, as we know, occurs at a speed of $v_S$$\approx\,$240~km/s. Therefore, the Newtonian galactic potential of attraction $U_\text{Galaxy}$ can be estimated as $|U_\text{Galaxy}|/c^2$$\sim v_S^2/c^2 $$\approx 0.8\times 10^{-6}$. As a result, we have a rough lower bound for the cosmological potential, $|\Phi|/c^2$$ >$$10^{-6}$. Moreover, since our Galaxy is only a small part of the Universe, we can expect a stronger condition, $|\Phi|/c^2\gg 10^{-6}$. Also, if we describe gravitational phenomena within the Solar system, then on such small scales we can absolutely neglect the spatial non-uniformity of the cosmological potential, i.e. $\Phi({\bf r})\approx const$.

Thus, we formulate the main approximation that we will use further:
\begin{equation}\label{main_approx}
|\nabla_{\bf r} U_\text{loc}({\bf r})|\gg |\nabla_{\bf r}\Phi({\bf r})|\,,
\end{equation}
where for the cosmological gravitational potential within the Solar System, the following lower bound holds:
\begin{equation}\label{CP_est}
|\Phi/c^2|> 10^{-6}\,,
\end{equation}
which was determined from the orbital velocity of the Sun around the center of the Galaxy. Moreover, summarizing the above numerical estimates, we have
\begin{equation}\label{3rd_approx}
|U_\text{loc}({\bf r})|\ll |\Phi({\bf r})|\,,
\end{equation}
at the distance of planetary orbits.

We add that the concept of finite cosmological potential is in agreement with the modern picture of origin and evolution of the Universe. Indeed, if the Universe (Metagalaxy) has been formed about 14 billion years ago as a result of the Big bang, then it is probably a finite space-time formation and, therefore, the cosmological gravitational potential is finite at each point of the Universe. Similar concept of cosmological potential [see expressions (\ref{U_structure})-(\ref{3rd_approx})] is universal and can be applied to most planetary systems with relatively small central stars.

\section{Post-Newtonian Parameters $\beta$ and $\gamma$}
In this section, we show that, from the viewpoint of the Laboratory Observer, the main post-Newtonian parameters $\beta$ and $\gamma$ depend on the cosmological potential $\Phi$. In the case of a small value, $|\Phi/c^2|$$\ll$1, these dependencies can be represented as:
\begin{eqnarray}\label{gamma_beta_gen}
&&\beta=\beta_0+\beta_1\Phi/c^2+O\left( \Phi^2/c^4\right),\nonumber\\
&&\gamma=\gamma_0+\gamma_1\Phi/c^2+O\left(\Phi^2/c^4\right),
\end{eqnarray}
where $\beta_0$ and $\gamma_0$ are PPN parameters from the viewpoint of the Cosmological Observer.

Our approach is based on the following scheme. We start with a cosmological description, in which we initially include some local system of small gravitating bodies (e.g., the Solar System) as a small part of more global system (e.g., Galaxy, galaxy cluster, Metagalaxy). Further, using the general expression for the metric tensor written in the framework of the PPN formalism, we consider a relatively small area of space for the analysis of gravitation only inside of our chosen system. This allows us to distinguish the cosmological background using the concept of cosmological potential [see Eqs.~(\ref{U_structure}) and (\ref{CP})], which is then reduced going to the reference frame of the local Laboratory Observer. Using the equation for the geodesic line, the result of such a reduction is expression (\ref{gamma_beta_gen}), which shows the difference between {\em measured} (within the Solar System) PPN parameters ($\beta,\gamma$) from their initial (cosmological) values ($\beta_0,\gamma_0$), i.e., there is a renormalization of the PPN parameters.

Note that this result can be explained (see Section~VI) by the non-invariance of the PPN formulation of general relativity with respect to the transformation $U({\bf r})\,\rightarrow\,U({\bf r})+C$, where $U({\bf r})$ is a gravitation potential and $C$ is an arbitrary real constant. In this case, the gravitational potential itself becomes really physical (in contrast to the classical Newton theory).

In order to accurately calculate the renormalized post-Newtonian parameters $\gamma$ and $\beta$ for the Laboratory Observer in the linear approximation with respect to the small cosmological parameter $|\Phi/c^2|$$\ll$1 [see Eq.~(\ref{gamma_beta_gen})], it is necessary to consider a general nonlinear post-post-Newtonian model of gravitation (see below).

\subsection*{\em IV.1 General nonlinear post-post-Newtonian model of gravitation in cosmological reference frame}
In this paper, we will use the PPN expression, which for the static distribution of masses has the form (see Refs.~\cite{Tur_2008,Einstein_1938})
\begin{align}\label{nonlinear_PN}
&g^{}_{00}({\bf r})=1+\frac{2U({\bf r})}{c^2}+\frac{2\beta_0U^2({\bf r})}{c^4}+\frac{2\xi_0 Q({\bf r})}{c^4}+\frac{2\kappa_0U^3({\bf r})}{c^6},\nonumber\\
&g^{}_{11}({\bf r})=1-\frac{2\gamma_0U({\bf r})}{c^2}+\frac{3\delta_0U^2({\bf r})}{2c^4},
\end{align}
which is the solution of gravitational fundamental equations (e.g., Einstein equations) in the reference frame of the Cosmological Observer corresponding to a plane cosmological space-time $\{t,{\bf r}\}_\text{Cosm}$. The potential $U({\bf r})$ is determined in accordance with formula (\ref{Newton_pot}), and the function $Q({\bf r})$ has the following form:
\begin{equation}\label{Q}
Q({\bf r})= \sum_{n}\frac{\Gamma_0 M_n}{|{\bf r}-{\bf R}_n|}\sum_{l\neq n}\frac{\Gamma_0 M_{l}}{|{\bf R}_n-{\bf R}_{l}|}\,.
\end{equation}
In expression (\ref{nonlinear_PN}), in addition to the standard post-Newtonian parameters $\gamma_0$ and $\beta_0$, there are two post-post-Newtonian parameters $\delta_0$ (see, for example, Ref.~\cite{Tur_2008}) and $\kappa_0$. It should be noted that the coefficient $\xi_0$ is not independent, but is expressed in terms of $\beta_0$: $\xi_0=(2\beta_0-1)$ (e.g., see Ref.~\cite{Tur_2008}). However, for simplicity, we will consider the coefficient $\xi_0$ as independent. For standard GR, there is $\beta_0=\gamma_0=\xi_0=\delta_0=1$, while information about $\kappa_0$ is not available in the scientific literature, to our knowledge.

Using the representation for the potential in the form (\ref{U_structure}), we now rewrite the expressions (\ref{nonlinear_PN}) up to the quadratic term $U^2_\text{loc}({\bf r})$ for the component $g^{}_{00}({\bf r})$ and up to the linear term $U^{}_\text{loc}({\bf r})$ for the component $g^{}_{11}({\bf r})$:
\begin{align}\label{nonlinear_PN_loc}
&g^{}_{00}({\bf r})=G_{00}+{\cal A}\frac{2U^{}_\text{loc}({\bf r})}{c^2}+{\cal B}\frac{2U^{2}_\text{loc}({\bf r})}{c^4}+\frac{2\xi_0 Q_\text{loc}({\bf r})}{c^4}\,,\nonumber\\
&g^{}_{11}({\bf r})=G_{11}-{\cal D}\frac{2U^{}_\text{loc}({\bf r})}{c^2}\,,
\end{align}
where we introduced a series of cosmological coefficients:
\begin{eqnarray}\label{ABD}
G_{00}&=& 1+\frac{2\Phi}{c^2}+\frac{2\xi_0 \Pi_\text{CP}}{c^4}+O\left(\frac{\Phi^2}{c^4}\right),\nonumber \\
G_{11}&=& 1-\gamma_0\frac{2\Phi}{c^2}+O\left(\frac{\Phi^2}{c^4}\right),\nonumber\\
{\cal A} &=& 1+(2\beta_0+\xi_0)\frac{\Phi}{c^2}+O\left(\frac{\Phi^2}{c^4}\right), \\
{\cal B} &=& \beta_0+3\kappa_0\frac{\Phi}{c^2}+O\left(\frac{\Phi^2}{c^4}\right),\nonumber\\
{\cal D} &=& \gamma_0-\frac{3\delta_0}{2}\frac{\Phi}{c^2}+O\left(\frac{\Phi^2}{c^4}\right).\nonumber
\end{eqnarray}
Here we used that the expression (\ref{Q}) for $Q({\bf r})$ can be represented with good accuracy in the following form (see Appendix~A):
\begin{equation}\label{Q2}
Q({\bf r})\approx U^{}_\text{loc}({\bf r})\Phi+\Pi_\text{CP}+Q_\text{loc}({\bf r})\,,
\end{equation}
where $\Pi_\text{CP}$ is another cosmological contribution that can also be considered as a constant within the local planetary system ($\Pi_\text{CP}=const$), and its value is estimated in order of magnitude as
\begin{equation}\label{Pi_approx}
\frac{\Pi_\text{CP}}{c^4}\ll \frac{|\Phi|}{c^2},
\end{equation}
and it is a contribution to the $G_{00}$ [see Eqs. (\ref{nonlinear_PN_loc}) and (\ref{ABD})]. The expression for $Q_\text{loc}({\bf r})$ in Eq.~(\ref{nonlinear_PN_loc}) has the form:
\begin{equation}\label{Q_loc}
Q_\text{loc}({\bf r})=\sum_{n=1}^{N_\text{loc}}\frac{\Gamma_0 M_n}{|{\bf r}-{\bf R}_n|}\sum_{l\neq n}^{N_\text{loc}}\frac{\Gamma_0 M_{l}}{|{\bf R}_n-{\bf R}_{l}|}\,,
\end{equation}
where, unlike Eq.~(\ref{Q}), the summation is only over the indexes $\{1,2, ..., N_\text{loc}\}$, which number the sources of gravitation only inside of the considered planetary system (the Solar System in our case).

\subsection*{\em IV.2 Metric components in laboratory reference frame}
Based on Einstein's principles, we introduce a reference frame associated with a local ``Laboratory Observer'', which, unlike the global and abstract ``Cosmological Observer'', is closely tied to real physical measurements in a relatively small area of the space. In our case, to describe gravitational phenomena within the Solar System, space-time for the Laboratory Observer is defined as follows. Neglecting the weak contribution of $U_\text{loc}({\bf r})$ in Eqs.~(\ref{nonlinear_PN_loc}), the expression for the interval has the form:
\begin{equation}\label{int_loc}
ds^2 =G_{00}(c\,dt)^2-G_{11} d{\bf r}^2\,.
\end{equation}
According to Einstein's basic postulate, a real Laboratory Observer is not able to detect the presence/absence of gravitation based on {\em strictly local} physical measurements (i.e., under constant gravitational potential). In particular, such local measurements include the measurement of the speed of light during its propagation between two arbitrarily close spatial points. Therefore, for the Laboratory Observer, the local speed of light in any region of the Universe should always be equal to $c$, regardless of the cosmological potential $\Phi({\bf r})$ for this region. Based on Einstein's postulate and the formula (\ref{int_loc}), we find the following transformation law:
\begin{equation}\label{lab_coor_2}
dt'=dt\sqrt{G_{00}}\;,\quad d{\bf r}'=d{\bf r}\sqrt{G_{11}}\;,
\end{equation}
which connects the space-time of the Laboratory Observer $\{t',{\bf r}'\}_\text{Lab}$ with the space-time of the Cosmological Observer $\{t,{\bf r}\}_\text{Cosm}$. In this case, the expression for the interval (\ref{int_loc}) in the reference frame $\{t',{\bf r}'\}_\text{Lab}$ takes the following form:
\begin{equation}\label{ds_Lab_free1}
ds^2 =(c\,dt')^2-(d{\bf r}')^2\,,
\end{equation}
i.e., laboratory space-time corresponds to the ideal Minkowski space-time, in which information about the cosmological potential of $\Phi$ is deeply ``hidden'' from the observer. Setting $ds=0$ in Eq.~(\ref{ds_Lab_free1}), we find that the speed of light for the Laboratory Observer is always equal to $c$ in any area of the Universe. Note also that the coordinates $\{t',{\bf r}'\}_\text{Lab}$, defined in Eq.~(\ref{lab_coor_2}), are named as ``local quasi-Cartesian coordinates'' in Ref.~\cite{Will_2018}.

Thus, the Laboratory Observer is a observer studying gravitation in a relatively small region of space near relatively small celestial bodies. In our case, we associate the reference frame of the Laboratory Observer with the spatial region of the Solar System, where the continuum $\{t',{\bf r}'\}_\text{Lab}$ is the flat space-time for a description of gravitational (and other physical) phenomena and experiments within the given system.

As a result of the transformations (\ref{lab_coor_2}), we obtain the following expression for the interval $ds^2$, rewritten in the variables of the laboratory reference frame $\{t',{\bf r}'\}_\text{Lab}$:
\begin{equation}\label{ds_Lab_free}
ds^2 =\tilde{g}_{00}({\bf r}')(c\,dt')^2-\tilde{g}_{11}({\bf r}')(d{\bf r}')^2\,,
\end{equation}
where, based on Eq.~(\ref{nonlinear_PN_loc}), we have:
\begin{align}\label{nonlinear_PN_lab}
&\tilde{g}_{00}({\bf r}')=1+\frac{2{\cal A}U^{}_\text{loc}({\bf r}')}{G_{00}\,c^2}+\frac{2{\cal B}U^{2}_\text{loc}({\bf r}')}{G_{00}\,c^4}+\frac{2\xi_0 Q_\text{loc}({\bf r}')}{G_{00}\,c^4}\,,\nonumber\\
&\tilde{g}_{11}({\bf r}')=1-\frac{2{\cal D}U^{}_\text{loc}({\bf r}')}{G_{11}\,c^2}\,.
\end{align}
Here the spatial dependence $U_\text{loc}({\bf r}')$, in accordance with the transformation law (\ref{lab_coor_2}), is defined as:
\begin{equation}\label{phi2}
U_\text{loc}({\bf r}')\equiv U_\text{loc}({\bf r})|_{{\bf r}\to{\bf r}'/\sqrt{G_{11}}}\,.
\end{equation}

\subsection*{\em IV.3 Gravitational potential and PPN parameters in laboratory reference frame}
However, in an experimental study of gravitation and testing of GR, the Laboratory Observer receives information about the gravitational potential not from purely theoretical calculations, but from experimental data on the motion of bodies under action of gravity. In particular, information on the Sun's gravitational potential is obtained from an analysis of the orbital motion of the planets and other small bodies. Thus, for the Laboratory Observer, the {\em experimental} determination of the gravitational potential $\varphi({\bf r}')$ is related to the equation of free motion in the framework of classical mechanics, which has the standard form:
\begin{equation}\label{mov_eq}
\frac{d^2 {\bf r}'}{d t'^2}=-\nabla_{{\bf r}'}\varphi({\bf r}')\,,
\end{equation}
where $\varphi({\bf r}')$ differs from the theoretical function $U_\text{loc}({\bf r}')$ defined in the framework of metric theory. Moreover, in addition to equation (\ref{mov_eq}), the gravitational potential of $\varphi({\bf r}')$ should also satisfy the additivity condition:
\begin{equation}\label{additiv}
\varphi({\bf r}')=\sum_{a}\varphi^{}_{a}({\bf r}'-{\bf R}'_{a})\,,
\end{equation}
where $\varphi^{}_{a}({\bf r}'-{\bf R}'_{a})$ is the gravitational contribution from the $a$-th point source with the coordinate ${\bf R}'_{a}$.

In order to determine the relationship between the functions $\varphi({\bf r}')$ and $U_\text{loc}({\bf r}')$, we write the equation for the geodesic line (\ref{geod_vec}) in the cosmological reference frame $\{t,{\bf r}\}_\text{Cosm}$ and in the linear approximation in $U_\text{loc}({\bf r})$:
\begin{equation}\label{geod_cosm}
\frac{d^2 {\bf r}}{d t^2}=-\frac{\cal A}{G_{11}}\nabla_{{\bf r}}U_\text{loc}({\bf r})\,,
\end{equation}
where we used the components of the metric tensor (\ref{nonlinear_PN_loc}) subject to the conditions (\ref{main_approx}). Rewriting equation (\ref{geod_cosm}) in the laboratory reference frame $\{t',{\bf r}'\}_\text{Lab}$ [i.e., using the linear transformations (\ref{lab_coor_2})], we obtain
\begin{equation}\label{geod_lab_2}
\frac{d^2 {\bf r}'}{d t'^2}=-\frac{\cal A}{G_{00}}\nabla_{{\bf r}'}U_\text{loc}({\bf r'})\,,
\end{equation}
where $\nabla_{{\bf r}'}$=${\bf e}_1(\partial /\partial x_1')$$+$${\bf e}_2(\partial /\partial x_2')$$+$${\bf e}_3(\partial /\partial x_3')$ is the spatial gradient operator in the laboratory reference frame $\{t',{\bf r}'\}_\text{Lab}$. Comparing Eq.~(\ref{geod_lab_2}) with the equation for free motion (\ref{mov_eq}) and the additivity condition (\ref{additiv}), we find the relationship between $\varphi({\bf r}')$ and $U_\text{loc}({\bf r}')$:
\begin{equation}\label{lab_cosm_2}
\varphi({\bf r}')=\frac{\cal A}{G_{00}}\,U_\text{loc}({\bf r'})\;\Leftrightarrow\;U_\text{loc}({\bf r'})=\frac{G_{00}}{\cal A}\,\varphi({\bf r}')\,.
\end{equation}
Note that the same result can be obtained immediately if, instead of equation (\ref{geod_vec}) in the cosmological reference frame $\{t,{\bf r}\}_\text{Cosm}$, we use the following equation for the geodesic line:
\begin{equation}\label{geod_lab_gen}
\frac{d^2 {\bf r}'}{d t'^2}=-\frac{c^2}{2\tilde{g}_{11}({\bf r}')}\nabla_{{\bf r}'}{\tilde{g}_{00}({\bf r}')}\,,
\end{equation}
written in the laboratory reference frame $\{t',{\bf r}'\}_\text{Lab}$.

On the one hand, the relationship (\ref{lab_cosm_2}) leads to the following form of the gravitational constant $\Gamma'$ for the Laboratory Observer (see Appendix~B):
\begin{equation}\label{Gamma}
\Gamma'=\frac{{\cal A}\sqrt{G_{11}}}{G_{00}}\,\Gamma_0\,,
\end{equation}
i.e., it differs from the initial value $\Gamma_0$ for the Cosmological Observer [see Eq.~(\ref{Newton_pot})] and depends on the cosmological gravitational background (via the coefficients ${\cal A}$, $G_{11} $, and $G_{00}$). On the other hand, Eq.~(\ref{lab_cosm_2}) allows us to express the components of the metric tensor (\ref{nonlinear_PN_lab}) in terms of the laboratory gravitational potential $\varphi({\bf r}')$:
\begin{eqnarray}\label{nonlinear_PN_lab_2}
&&\tilde{g}_{00}({\bf r}')=1+\frac{2\varphi({\bf r}')}{c^2}+\frac{2\beta\varphi^2({\bf r}')}{c^4}+\frac{2\xi q({\bf r}')}{c^4}\,,\nonumber\\
&&\tilde{g}_{11}({\bf r}')=1-\frac{2\gamma\varphi({\bf r}')}{c^2}\,,
\end{eqnarray}
where, using Eq.~(\ref{Gamma}), the functions $\varphi({\bf r}')$ and $q({\bf r}')$ can be written as
\begin{eqnarray}\label{phi_q}
\varphi({\bf r}') &=& -\sum_{n=1}^{N_\text{loc}}\frac{\Gamma' M_n}{|{\bf r}'-{\bf R}'_n|}\,, \\
q({\bf r}') &=& \sum_{n=1}^{N_\text{loc}}\frac{\Gamma' M_n}{|{\bf r}'-{\bf R}'_n|}\sum_{l\neq n}^{N_\text{loc}}\frac{\Gamma' M_{l}}{|{\bf R}'_n-{\bf R}'_{l}|}\,.
\end{eqnarray}
The post-Newtonian parameters $\beta$, $\gamma$, and $\xi$ in Eq.~(\ref{nonlinear_PN_lab_2}) from the viewpoint of the Laboratory Observer are defined as follows:
\begin{equation}\label{beta_gamma}
\beta=\frac{G_{00}{\cal B}}{{\cal A}^2}\,,\quad \gamma=\frac{G_{00}{\cal D}}{G_{11}{\cal A}}\,,\quad \xi=\frac{G_{00}\xi_0}{{\cal A}^2}\,.
\end{equation}
It is these quantities that can be measured in real experiments, rather than the initial $\beta_0$, $\gamma_0$, and $\xi_0$ [see in Eq.~(\ref{nonlinear_PN})] that appear in the theoretical description of gravitation from the viewpoint of the virtual Cosmological Observer.

Thus, we have shown the principal possibility of cosmological gravimetry, information about which is contained in the post-Newtonian parameters $\beta$, $\gamma$ and $\xi$.
Note that our approach leads to different results compared to the well-known PPN approach presented in Ref.~\cite{Will_2018}, which also takes into account the cosmological gravitational background for local system of small gravitating bodies. A detailed comparison of both approaches and corresponding conclusions are presented in Appendix C, where we show the self-contradictoriness of the approach \cite{Will_2018}.

\subsection*{\em IV.4 Phenomenological cosmological generalization of post-Newtonian model}
Note that the above results were obtained in the model for a weak cosmological potential, $|\Phi/c^2|\ll 1$. However, expressions (\ref{nonlinear_PN_loc}) and (\ref{lab_coor_2})-(\ref{beta_gamma}) can be considered in a more general context, as a universal phenomenological generalization independent of the initial equations for $g_{jk}$. In this case, the parameters $G^{}_{00}$ and $G^{}_{11} $ should be interpreted as components of the cosmological metric tensor $G^\text{(Cosm)}_{jk}$:
\begin{equation}\label{G_Cosm}
G^\text{(Cosm)}_{jk}=\left(
  \begin{array}{cccc}
    G^{}_{00} & 0 & 0 & 0 \\
    0 & -G^{}_{11} & 0 & 0 \\
    0 & 0 & -G^{}_{11} & 0 \\
    0 & 0 & 0 & -G^{}_{11} \\
  \end{array}
\right),
\end{equation}
which describes the Universe as a whole and is the subject of research of various cosmological models, including models with ``dark'' energy and matter, and even models that go beyond the standard Einstein equations (see, for example, the review \cite{Will_2014}). Other cosmological parameters, ${\cal A}$, ${\cal B}$, and ${\cal D}$, are due to the nonlinearity of the initial equations for $g_{jk}$ in the cosmological reference frame. Moreover, from a cosmological viewpoint, all the specified parameters \{$G^{}_{00}(t,{\bf r})$, $G^{}_{11}(t,{\bf r})$, ${\cal A}(t,{\bf r})$, ${\cal B}(t,{\bf r})$, ${\cal D}(t,{\bf r})$\} depend on coordinates (for example, due to the inhomogeneous distribution of matter in the Universe) and time (due to the general cosmic evolution, as well as changes in the mutual spatial arrangement of star systems and neighboring galaxies). However, within the local planetary system these parameters can be considered as constants, and their values are determined by the spatio-temporal location of this planetary system (for example, the Solar System in our case) within the framework of the general cosmological picture of the evolving Universe in general, and our Galaxy in particular.

\section{Cosmological Gravimetry in Einstein's General Relativity}
If we now substitute the expressions for coefficients (\ref{ABD}) into Eq.~(\ref{beta_gamma}), then, up to a linear contribution of $\Phi/c^2$, we write the expressions for $\beta$:
\begin{eqnarray}\label{beta_PN}
&&\beta=\beta_0+\beta_1\frac{\Phi}{c^2}+O\left(\frac{\Phi^2}{c^4}\right);\nonumber\\
&&\beta_1=2\beta_0(1-2\beta_0-\xi_0)+3\kappa_0\,,
\end{eqnarray}
and for $\gamma$:
\begin{eqnarray}\label{gamma_PN}
&&\gamma=\gamma_0+\gamma_1\frac{\Phi}{c^2}+O\left(\frac{\Phi^2}{c^4}\right);\nonumber\\
&&\gamma_1=\gamma_0(2\gamma_0+2-2\beta_0-\xi_0)-3\delta_0/2\,.
\end{eqnarray}
In the case of Einstein's general relativity, $\beta_0=\gamma_0=\xi_0=\delta_0=1$, we obtain:
\begin{equation}\label{beta_GR}
\beta -1 = (3\kappa_0-4)\frac{\Phi}{c^2}+O\left(\frac{\Phi^2}{c^4}\right),
\end{equation}
\begin{equation}\label{gamma_GR}
\gamma -1 = -0.5\frac{\Phi}{c^2}+O\left(\frac{\Phi^2}{c^4}\right).
\end{equation}
Taking into account the estimate (\ref{CP_est}) for the cosmological potential $\Phi$, we can give a general estimate for the experimentally measured post-Newtonian parameter $\gamma$:
\begin{equation}\label{gamma_est}
|\gamma -1|> 0.5\times10^{-6},
\end{equation}
which should be fulfilled if Einstein's general relativity is true. Moreover, because $\Phi^{}_\text{CP}<0$ (as an usual Newtonian potential) in Eq.~(\ref{gamma_GR}), we can expect the following inequality
\begin{equation}\label{inequality_GR}
\gamma -1 >0\,,
\end{equation}
which can be considered as an additional test of standard GR and cosmology. However, if experiments will show $(\gamma -1)<0$, then this can be interpreted as the existence of antigravitation as the dominant cosmological factor (i.e., $\Phi^{}_\text{CP}>0$).

Because the value $\kappa_0$ for GR is still unknown, it is impossible to draw a final conclusion about the post-Newtonian coefficient $\beta$ in (\ref{beta_GR}). However, we believe with high probability that for Einstein's theory an estimate similar to (\ref{gamma_est}) is fulfilled:
\begin{equation}\label{beta_est}
|\beta -1|> 10^{-6}.
\end{equation}
Moreover, if the condition $|\Phi/c^2|\ll 1$ is satisfied, then Eqs.~(\ref{beta_PN}) and (\ref{gamma_PN}) allow us to formulate a cosmological test for any post-Newtonian theory of gravitation:
\begin{equation}\label{test_PN}
\frac{\beta-\beta_0}{\gamma-\gamma_0}\approx\frac{\beta_1}{\gamma_1}\,,
\end{equation}
which, in the case of Einstein's theory, has the form:
\begin{equation}\label{test_GR}
\frac{\beta-1}{\gamma-1}\approx 2(4-3\kappa_0)\,.
\end{equation}
Thus, the magnitude of this ratio can be unambiguous cosmological test for standard GR.

Note that the above results were obtained in the framework of the mathematical model (\ref{nonlinear_PN}), where we took into account only two post-post-Newtonian contributions: the contribution $U^3({\bf r})/c^6$ for the component $g_{00}$, and the contribution $U^2({\bf r})/c^4 $ for the component $g_{11} $. In this case, the possible existence of other post-post-Newtonian contributions can lead to some correction of the expressions for $\beta_1$ and $\gamma_1$ in (\ref{beta_PN})-(\ref{gamma_PN}), and therefore for the relations (\ref{beta_GR})-(\ref{test_GR}). Thus, to clarify the presented results, we need to obtain complete expressions for all the main post-post-Newtonian contributions, which are absent in the scientific literature, to our knowledge. Such contributions include: all contributions proportional to $c^{-6}$ for the metric component $g_{00}$ and all contributions proportional to $c^{-4}$ for the metric component $g_{11}$ [see Eq.~(\ref {nonlinear_PN})]. Nevertheless, the above analysis clearly demonstrates the principal possibility of cosmological gravimetry both within the framework of any metric theory of gravitation in general and within the framework of the standard general relativity in particular.

\section{Cosmologically Insensitive Metric Model of Gravitation}
In the previous sections, we have shown the possibility of cosmological gravimetry in the framework of metric theories of gravitation, including Einstein's general relativity. The essence of this cosmological gravimetry is that the gravitational potential of the entire Universe $\Phi$, despite its highest degree of spatial homogeneity on small scales, can nevertheless have a certain effect on some gravitational measurements carried out within a relatively local system of small gravitating bodies (the Solar System, in our case). This allows us to measure (at least in principle) the value of $\Phi$ for a given region of space. However, as will be shown below, there is at least one distinguished metric model in which the possibility of such cosmological gravimetry is absent.

Abstracting from Einstein equations, we will build a nonlinear PPN model that satisfies the following three postulates:

1. Gravitation is described by the metric tensor $g_{jk}$ [see Eq.~(\ref{int})].

2. In the case of a small gravitational potential, $|U({\bf r})/c^2|\ll 1$, the model should coincide with the classical Newtonian theory of gravitation.

3. The model should not allow us to measure the cosmological potential $\Phi$.

These postulates are satisfied by a model with a diagonal metric tensor (\ref{g_gen}), whose components have the following form:
\begin{equation}\label{g_noCG}
g_{00}({\bf r})=e^{2U({\bf r})/c^2},\quad g_{11}({\bf r})=e^{-2U({\bf r})/c^2}.
\end{equation}
Indeed, taking into account the cosmological potential $\Phi$ in the expression (\ref{U_structure}), we rewrite the metric components (\ref{g_noCG}) as:
\begin{align}\label{g_noCG_struct}
&g_{00}({\bf r})=G_{00\,}e^{2U_\text{loc}({\bf r})/c^2},\quad G_{00}=e^{2\Phi({\bf r})/c^2},\nonumber \\
&g_{11}({\bf r})=G_{11\,}e^{-2U_\text{loc}({\bf r})/c^2},\quad G_{11}=e^{-2\Phi({\bf r})/c^2},
\end{align}
where $G_{00}$ and $G_{11}$ should be considered as cosmological components. Using the transformations (\ref{lab_coor_2}), we go to the reference frame of the Laboratory Observer $\{t',{\bf r}'\}_\text {Lab}$ with the metric tensor (\ref{ds_Lab_free}). In this reference frame, using Eq.~(\ref{g_noCG_struct}), we obtain:
\begin{equation}\label{g_noCG_loc}
\tilde{g}_{00}({\bf r}')=e^{2U_\text{loc}({\bf r}')/c^2},\quad \tilde{g}_{11}({\bf r}')=e^{-2U_\text{loc}({\bf r}')/c^2},
\end{equation}
where the spatial dependence $U_\text{loc}({\bf r}')$ is determined in accordance with (\ref{phi2}).

Using expressions (\ref{g_noCG_loc}), we write the equation for the geodesic line (\ref{geod_lab_gen}) in the laboratory reference frame $\{t',{\bf r}'\}_\text{Lab}$:
\begin{equation}\label{geod_lab_noCG}
\frac{d^2 {\bf r}'}{d t'^2}=-\nabla_{{\bf r}'}U_\text{loc}({\bf r'})\,,
\end{equation}
where we restricted ourselves to a linear contribution of $U_\text{loc}({\bf r'})$. Comparing this equation with the equation for free motion (\ref{mov_eq}), we find the relationship between $U_\text{loc}({\bf r'})$ and the gravitational potential $\varphi({\bf r}')$ from the viewpoint of the Laboratory Observer:
\begin{equation}\label{lab_cosm_noCG}
\varphi({\bf r}')=U_\text{loc}({\bf r'})\,.
\end{equation}
As a result, we obtain the final expressions:
\begin{equation}\label{g_noCG_Lab}
\tilde{g}_{00}({\bf r}')=e^{2\varphi({\bf r}')/c^2},\quad \tilde{g}_{11}({\bf r}')=e^{-2\varphi({\bf r}')/c^2},
\end{equation}
in which there is no ``trace'' (information) of the cosmological potential $\Phi$. Moreover, in the case of $|\varphi({\bf r}')/c^2|\ll 1$, the decomposition takes place:
\begin{align}\label{g_noCG_composition}
&\tilde{g}_{00}({\bf r}')=1+\frac{2\varphi({\bf r}')}{c^2}+\frac{2\varphi^2({\bf r}')}{c^4}+\frac{4\varphi^3({\bf r}')}{3c^6}+...\,,\nonumber \\
&\tilde{g}_{11}({\bf r}')=1-\frac{2\varphi({\bf r}')}{c^2}+\frac{2\varphi^2({\bf r}')}{c^4}+... \,,
\end{align}
from which Newton's theory for weak gravitational fields follows [see the main (Newtonian) term, $2\varphi({\bf r}')/c^2$, in the expression for $\tilde{g}_{00}({\bf r}')$].

Thus, we have proved that the metric model (\ref{g_noCG}) fully satisfies Postulates 1-3. It is also obvious that Postulate 3 describes, in fact, the invariance of the theory under the transformation:
\begin{equation}\label{invar}
U({\bf r})\,\rightarrow\,U({\bf r})+C_0\,,
\end{equation}
where $C_0$ is an arbitrary real constant. In this case, the value of the gravitational potential itself is not really physical (as in Newton's theory of gravitation). In essence, it is precisely the non-invariance with respect to the transformation (\ref{invar}) that makes cosmological gravimetry possible for other metric models (including Einstein's theory) for which the value of the gravitational potential itself becomes really physical.

Moreover, from expressions (\ref{g_noCG}), (\ref{g_noCG_Lab}) and (\ref{g_noCG_composition}) we find the values of the post-Newtonian and post-post-Newtonian parameters [see in Eqs.~(\ref{nonlinear_PN}) and (\ref{nonlinear_PN_lab_2})]:
\begin{eqnarray}\label{beta_gamma_noCG}
 && \beta=\beta_0=\gamma=\gamma_0=1\,,\quad \xi=\xi_0=0\,,\\
 && \delta=\delta_0=4/3\,, \quad \kappa=\kappa_0=2/3\,,\nonumber
\end{eqnarray}
which are the same for both Cosmological and Laboratory Observers. It follows that all the currently known successful experimental tests of GR within the Solar System [such as: measuring the anomalous perihelion shift of Mercury's orbit, measuring the deflection of light near the Sun, the time delay of radar signals sent across the Solar System past the Sun to a planet or satellite and returned to the Earth (due to the gravitation of the Sun), laser ranging of the Moon, etc.] can also be considered, from a formal viewpoint, as confirmation of the cosmologically insensitive model of gravitation presented above, because all these experiments show: $\beta\approx 1$ and $\gamma\approx 1$. In this context, the experimental detection of deviations, $|\beta-1|>10^{-6}$ and $|\gamma-1|>10^{-6}$, will not only be a demonstration of cosmological gravimetry, but it will also be a critical test for Einstein's general relativity. Indeed, if experiments will show $|\beta-1|\ll 10^{-6}$ and $|\gamma-1|\ll 10^{-6}$, then the cosmologically insensitive model of gravitation presented above will be more preferred.

We stress that the requirement of the invariance with respect to transformation (\ref{invar}) (or Postulate~3) is so very strong that it allows us to find the solution (\ref{g_noCG}) without appeal to any equations for the metric tensor $g_{jk}$, which are unknown in this case. Therefore, a special theoretical interest may be the problem of constructing a more complete and closed theory, in which the expressions (\ref{g_noCG}) are the solution of some basic equations (at least for the static distribution of masses). Some interesting and unusual features of the cosmologically insensitive model of gravitation are presented in Appendix~D.

\subsection*{6. Discussion and conclusion}
As shown above, the key factor for the possibility of cosmological gravimetry is the non-invariance of the theory with respect to transformation (\ref{invar}). Indeed, if some theory is non-invariant with respect to Eq.~(\ref{invar}), then in such a theory the gravitational potential is a completely definite (fixed) value, i.e., the gravitational potential itself becomes really physical \cite{we}. This, in turn, allows us also to formulate the concept of the cosmological gravitational potential $\Phi $ as a well-defined physical value and raises the question of its experimental measurement. This circumstance contrasts with the classical linear theory of gravitation based on the gravitational force ${\bf F}_\text{grav}=-M\nabla U({\bf r})$, which leads to the invariance with respect to transformation (\ref{invar}) and therefore to the impossibility of a cosmological gravimetry.

Note that an idea of the possibility of cosmological gravimetry in the case of non-invariance with respect to transformation (\ref{invar}) was first proposed in Ref.~\cite{Yudin_2019}, where the gravitational redshift in atomic clocks was considered in the framework of the nonmetric description (due to the mass defect effect in quantum atomic physics, e.g., see Refs.~\cite{Yudin_2018,Wolf_2016}). However, in our paper, we have substantiated the possibility of a cosmological gravimetry for the metric nonlinear theories of gravitation, including Einstein's general relativity. Such a gravimetry is based on the experimental measurement of two main post-Newtonian parameters, the values of which are different for the Laboratory Observer ($\beta,\gamma $) and for the Cosmological Observer ($\beta_0,\gamma_0 $). For instance, in the case of GR, information on cosmological gravitation is contained in the values $(\beta-1)$ and $(\gamma-1)$. Moreover, a lower estimate was obtained for the Solar System: $|\beta-1|>10^{-6}$ and $|\gamma-1|>10^{-6} $, based on the velocity of the Sun's orbital motion around the Galaxy center. Also, a cosmological test for GR is formulated, in which the ratio $(\beta-1)/(\gamma-1)$ has a quite certain value, which can be analytically calculated if all post-Newtonian and post-post-Newtonian contributions in the metric tensor are known (this work still needs to be completed). The relationship between the gravitational constant $\Gamma'$ for the laboratory reference frame and its initial value $\Gamma_0$ in the cosmological reference frame is also shown.

Note that we used a simplified model of the motionless Solar System under static distribution of masses in the Universe. Therefore, the Sun's orbital motion and the motion of celestial bodies can, in principle, significantly affect the presented results if the cosmological potential $\Phi$ does not far exceed the gravitational potential of our Galaxy (i.e., if $\Phi\approx U_\text{Galaxy}$). However, in any case, this does not negate the concept of cosmological gravimetry developed here.

Moreover, a unique PPN model was constructed, which is invariant with respect to transformation (\ref{invar}) and therefore it has no possibility for cosmological gravimetry. The most unusual is that for this model we have, $\beta=\beta_0=\gamma=\gamma_0=1$, which is very close to Einstein's theory. Therefore, the experimental detection of deviations, $|\beta-1|>10^{-6}$ and $|\gamma-1|>10^{-6}$, becomes critical for GR. Indeed, if experiments will show $|\beta-1|\ll 10^{-6}$ and $|\gamma-1|\ll 10^{-6}$, then the cosmologically insensitive model of gravitation will be more preferred.

In the context of the above, it can be argued that all experiments on measuring post-Newtonian parameters ($\beta,\gamma$) should be considered as experimental cosmological gravimetry. Such gravimetry, in the case of deviations of $|\beta-1|>10^{-6}$ and $|\gamma-1|>10^{-6}$, will open up new unique opportunities for investigation of the Universe and will be additional test of Einstein's general relativity. At the same time, we note that the indicated level of accuracy ($10^{-6}$) is quite achievable using modern measuring equipment. Therefore, the obtained results can be used as additional motivation for new space-based experiments with the aim of more accurate measurement of the PPN parameters ($\beta,\gamma$).


\appendix

\section{}
We represent expression (\ref{Q}) in the form of the following three contributions:
\begin{align}\label{Q_superpos}
& Q({\bf r})=Q^{}_1({\bf r})+\Pi({\bf r})+Q_\text{loc}({\bf r})\,, \\
& Q^{}_1({\bf r})=\sum_{n=1}^{N_\text{loc}}\frac{\Gamma_0 M_n}{|{\bf r}-{\bf R}_n|}\sum_{l>N_\text{loc}}\frac{\Gamma_0 M_{l}}{|{\bf R}_n-{\bf R}_{l}|}\,,\nonumber\\
& \Pi({\bf r})=\sum_{n>N_\text{loc}}\frac{\Gamma_0 M_n}{|{\bf r}-{\bf R}_n|}\sum_{l\neq n}\frac{\Gamma_0 M_{l}}{|{\bf R}_n-{\bf R}_{l}|}\,,\nonumber \\
& Q_\text{loc}({\bf r})=\sum_{n=1}^{N_\text{loc}}\frac{\Gamma_0 M_n}{|{\bf r}-{\bf R}_n|}\sum_{l\neq n}^{N_\text{loc}}\frac{\Gamma_0 M_{l}}{|{\bf R}_n-{\bf R}_{l}|}\,.\nonumber
\end{align}
The first term $Q^{}_1({\bf r})$ can be written as:
\begin{align}\label{Q1}
Q^{}_1({\bf r})=&\sum_{n=1}^{N_\text{loc}}\frac{-\Gamma_0 M_n}{|{\bf r}-{\bf R}_n|}\sum_{l>N_\text{loc}}\frac{-\Gamma_0 M_{l}}{|{\bf R}_n-{\bf R}_{l}|}= \nonumber\\
&\sum_{n=1}^{N_\text{loc}}\frac{-\Gamma_0 M_n \Phi({\bf R}_n)}{|{\bf r}-{\bf R}_n|}\,,
\end{align}
where $\Phi({\bf R}_n)$ is the cosmological potential [see Eq.~(\ref{CP})] at the point where the gravitation source ${\bf R}_n$ is inside of our local system of celestial bodies (i.e., the Solar System). However, because the cosmological potential is almost uniform within the studied local system [$\Phi({\bf r})\approx const$], then, using (\ref{U_local}), we can write:
\begin{equation}\label{Q1_approx}
Q_1({\bf r})\approx\Phi\sum_{n=1}^{N_\text{loc}}\frac{-\Gamma_0 M_n}{|{\bf r}-{\bf R}_n|}=U_\text{loc}({\bf r})\Phi\,.
\end{equation}
The second term in (\ref{Q_superpos}) can be represented as:
\begin{align}\label{Pi_sup}
&\Pi({\bf r})=\sum_{n>N_\text{loc}}\frac{-\Gamma_0 M_n}{|{\bf r}-{\bf R}_n|}\sum_{l\neq n}\frac{-\Gamma_0 M_{l}}{|{\bf R}_n-{\bf R}_{l}|}= \\
& \sum_{n>N_\text{loc}}\frac{-\Gamma_0 M_n \Phi({\bf R}_n)}{|{\bf r}-{\bf R}_n|}=\sum_{n>N_\text{loc}}\frac{\Gamma_0 M_n |\Phi({\bf R}_n)|}{|{\bf r}-{\bf R}_n|}\,. \nonumber
\end{align}
Because only deep space bodies ($n>N_\text{loc}$) are presented here, $\Pi({\bf r})$ should be considered as an additional cosmological contribution, which is spatially homogeneous within the considered local system, $\Pi({\bf r})\approx const$. To estimate the value of $\Pi({\bf r})$, we divide expression (\ref{Pi_sup}) by $c^2 $, that leads to the following inequality:
\begin{align}\label{Pi_est}
\frac{\Pi({\bf r})}{c^2}=&\sum_{n>N_\text{loc}}\frac{\Gamma_0 M_n |\Phi({\bf R}_n)|}{|{\bf r}-{\bf R}_n|\,c^2}\ll \nonumber\\
&\sum_{n>N_\text{loc}}\frac{\Gamma_0 M_n}{|{\bf r}-{\bf R}_n|}=|\Phi({\bf r})|\,,
\end{align}
where we used the condition $|\Phi({\bf R}_n)/c^2|\ll 1$. Going to dimensionless quantities, the inequality (\ref{Pi_est}) can be rewritten as:
\begin{equation}\label{Pi_est2}
\frac{\Pi({\bf r})}{c^4}\ll \frac{|\Phi({\bf r})|}{c^2}\,.
\end{equation}
The expression for $Q_\text{loc}({\bf r})$ in Eq.~(\ref{Q_superpos}) is not changed.

Thus, we finally write:
\begin{equation}\label{Q_approx}
Q({\bf r})\approx U_\text{loc}({\bf r})\Phi+\Pi({\bf r})+Q_\text{loc}({\bf r})\,,
\end{equation}
where for the cosmological contribution $\Pi({\bf r})$ the estimate (\ref{Pi_est2}) holds.

\section{}
Let us consider the gravitational potential created by the point mass $M$ with the coordinate ${\bf R}$ in the cosmological reference frame $\{t,{\bf r}\}_\text{Cosm}$:
\begin{equation}\label{U_cosm}
U({\bf r})=-\frac{\Gamma_0 M}{|{\bf r}-{\bf R}|}\,.
\end{equation}
Then, taking into account definition (\ref{phi2}), we find:
\begin{equation}\label{U_cosm_lab}
U({\bf r}')=-\frac{\Gamma_0 M \sqrt{G_{11}}}{|{\bf r}'-{\bf R}'|}\,.
\end{equation}
At the same time, the Newtonian gravitational potential $\varphi({\bf r}')$ for the Laboratory Observer has the standard form:
\begin{equation}\label{phi_lab}
\varphi({\bf r}')=-\frac{\Gamma' M'}{|{\bf r}'-{\bf R}'|}\,,
\end{equation}
where $\Gamma'$ and $M'$ are the gravitational constant and mass, respectively, in the reference frame $\{t',{\bf r}'\}_\text{Lab}$. Substituting now expressions (\ref{U_cosm_lab}) and (\ref{phi_lab}) in Eq.~(\ref{lab_cosm_2}), we obtain:
\begin{equation}\label{phi_lab_Ulab}
\frac{\Gamma' M'}{|{\bf r}'-{\bf R}'|}=\frac{\cal A}{G_{00}}\,\frac{\Gamma_0 M \sqrt{G_{11}}}{|{\bf r}'-{\bf R}'|}\,,
\end{equation}
that leads to the transformation law:
\begin{equation}\label{Gamma_M}
\Gamma' M'=\frac{{\cal A}\sqrt{G_{11}}}{G_{00}}\,\Gamma_0 M \,.
\end{equation}
If we assume that the rest masses are equal for any reference systems ($M=M'$), then we finally write the following transformation law for the gravitational constant:
\begin{equation}\label{Gamma_Lab}
\Gamma' =\frac{{\cal A}\sqrt{G_{11}}}{G_{00}}\,\Gamma_0 \,.
\end{equation}
Using the expressions from Eq.~(\ref{ABD}), we find:
\begin{equation}\label{Gamma_est}
\frac{\Gamma'}{\Gamma_0} =1+(2\beta_0-\gamma_0+\xi_0-2)\frac{\Phi}{c^2}+O(\Phi^2/c^4) \,.
\end{equation}
In the case of GR ($\beta_0=\gamma_0=\xi_0=1$), there is:
\begin{equation}\label{Gamma_GR}
\frac{\Gamma'}{\Gamma_0} =1+O(\Phi^2/c^4) \,.
\end{equation}
which means relatively weak dependence on cosmological gravitational background (at least, in the case of  $|\Phi/c^2|\ll 1$).

\section{}
Let us consider the well-known approach, which describes the general method of calculating post-Newtonian limits of metric theories of gravity taking into account the cosmological gravitational background (e.g., see {\em Section}~5.1 in Ref.~\cite{Will_2018}). Here we will reproduce this approach using our notations and the model of static distribution of masses. For simplicity, we will consider the spatial region near single point mass $M$, located far from other celestial bodies. In this case, the local gravitational potential in the cosmological reference frame $\{t,{\bf r}\}_\text{Cosm}$ has the form
\begin{equation}\label{App_U_CO}
U({\bf r})=-\frac{\Gamma_0 M}{|{\bf r}-{\bf r}_0|}\,,
\end{equation}
where ${\bf r}_0$ is the coordinate of the point mass $M$.

At the initial stage, Ref.~\cite{Will_2018} considers a metric in the cosmological reference frame $\{t,{\bf r}\}_\text{Cosm}$ taking into account only the linear contribution in $U({\bf r})$. According to our expressions (\ref{nonlinear_PN_loc}) and (\ref{App_U_CO}), this can be represented as follows:
\begin{equation}\label{App_ds_CO}
ds^2=\left(G^{}_{00}-\frac{2{\cal A}\Gamma_0 M}{c^2|{\bf r}-{\bf r}_0|}\right)(c\,dt)^2-G^{}_{11}(d{\bf r})^2\,.
\end{equation}
Further, using the law of transformations (\ref{lab_coor_2}) for Eq.~(\ref{App_ds_CO}), we obtain the expression
\begin{eqnarray}\label{App_ds_LO}
ds^2&=&\left(1-\frac{2{\cal A}\sqrt{G^{}_{11}}\,\Gamma_0 M}{G^{}_{00}c^2|{\bf r}'-{\bf r}'_0|}\right)(c\,dt')^2-(d{\bf r}')^2=\nonumber \\
&&\left(1-\frac{2\Gamma' M}{c^2|{\bf r}'-{\bf r}'_0|}\right)(c\,dt')^2-(d{\bf r}')^2\,,
\end{eqnarray}
in the laboratory reference frame $\{t',{\bf r}'\}_\text{Lab}$ (which is named as ``local quasi-Cartesian coordinates'' in Ref.~\cite{Will_2018}). In this case, the value
\begin{equation}\label{App_Gamma_LO}
\Gamma' =\frac{{\cal A}\sqrt{G^{}_{11}}}{G^{}_{00}}\,\Gamma_0
\end{equation}
should be considered as the renormalized gravitational constant in the laboratory reference system (see Eq.~(5.9) in Ref.~\cite{Will_2018}, where we need to make a replacement: $\alpha$$\to$${\cal A}$, $c_0$$\to$$G^{}_{00}$, $c_1$$\to$$G^{}_{11}$). As we see, Eq.~(\ref{App_Gamma_LO}) coincides with our formula (\ref{Gamma}).

At the final stage, in order to write the post-Newtonian terms in the laboratory reference frame $\{t',{\bf r}'\}_\text{Lab}$, Ref.~\cite{Will_2018} proposes to solve the Einstein equations for the point mass $M$ (in our case) in empty space (i.e., already without a cosmological background), but with the renormalized gravitational constant $\Gamma'$ from Eq.~(\ref{App_Gamma_LO}). Note that almost all expressions in Ref.~\cite{Will_2018} are represented using the ``conserved'' density $\rho^{\ast}$, which includes either the gravitational constant $\Gamma_0$ for the cosmological reference frame $\{t,{\bf r}\}_\text{Cosm}$, or $\Gamma'$ for ``local quasi-Cartesian coordinates'' $\{t',{\bf r}'\}_\text{Lab}$. Thus, following to Ref.~\cite{Will_2018}, in the laboratory reference frame  $\{t',{\bf r}'\}_\text{Lab}$, we obtain the following expression:
\begin{eqnarray}\label{App_ds_LO_GR}
ds^2&=&\left(1-\frac{2\Gamma' M}{c^2|{\bf r}'-{\bf r}'_0|}+\frac{2(\Gamma' M)^2}{c^4|{\bf r}'-{\bf r}'_0|^2}\right)(c\,dt')^2\nonumber \\
&&-\left(1+\frac{2\Gamma' M}{c^2|{\bf r}'-{\bf r}'_0|}\right)(d{\bf r}')^2\,,
\end{eqnarray}
which corresponds to the post-Newtonian coefficients $\beta=\gamma=1$ and, consequently, to the lack of the possibility of a cosmological gravimetry.

However, this final step (i.e., the use of the Einstein equations with the renormalized gravitational constant $\Gamma'$ for the local system of small gravitating bodies in the laboratory ``empty Universe'') does not follow from the basic principles of GR and it can be considered as intuitively-phenomenological approach (artificial, in our opinion), which allows in fact to take not into account the cosmological gravitational background when analyzing gravitational phenomena in the local planetary system. Below we show that the expression (\ref{App_ds_LO_GR}) leads to a contradiction. To do this, we use the transformation law (\ref{lab_coor_2}) to make in Eq.~(\ref{App_ds_LO_GR}) a reverse transition to the cosmological reference frame $\{t,{\bf r}\}_\text{Cosm}$. As a result, we get the following expression:
\begin{align}\label{App_ds_CO_GR}
ds^2&=g^{}_{00}({\bf r})(c\,dt)^2-g^{}_{11}({\bf r})(d{\bf r})^2\nonumber \\
&=\left(G^{}_{00}-\frac{2G^{}_{00}\Gamma' M}{\sqrt{G^{}_{11}}c^2|{\bf r}-{\bf r}_0|}+\frac{2G^{}_{00}(\Gamma' M)^2}{G^{}_{11}c^4|{\bf r}-{\bf r}_0|^2}\right)(c\,dt)^2\nonumber \\
&-\left(G^{}_{11}+\frac{2\sqrt{G^{}_{11}}\,\Gamma' M}{c^2|{\bf r}-{\bf r}_0|}\right)(d{\bf r})^2\,.
\end{align}
For example, let us consider in (\ref{App_ds_CO_GR}) the spatial metric component $g^{}_{11}({\bf r})$:
\begin{align}\label{App_g11_Will}
g^{}_{11}({\bf r})=\left(G^{}_{11}+\frac{{\cal A}G^{}_{11}}{G^{}_{00}}\frac{2\Gamma_0 M}{c^2|{\bf r}-{\bf r}_0|}\right),
\end{align}
where we have used the expression (\ref{App_Gamma_LO}) for $\Gamma'$. On the other hand, we have the {\em basic} expression (\ref{nonlinear_PN_loc}) for metric components obtained directly from solving of the Einstein equations in the cosmological reference frame $\{t,{\bf r}\}_\text{Cosm}$ with taking into account post-Newtonian and post-post-Newtonian terms [see Eq.~(\ref{nonlinear_PN})] and without additional artificial assumptions. In this case, the formula for the spatial metric component $g^{}_{11}({\bf r})$ has the form:
\begin{equation}\label{App_g11_CO}
g^{}_{11}({\bf r})=\left(G^{}_{11}+{\cal D}\frac{2\Gamma_0 M}{c^2|{\bf r}-{\bf r}_0|}\right),
\end{equation}
in which we use the local potential (\ref{App_U_CO}).

Now we will compare Eq.~(\ref{App_g11_Will}) and Eq.~(\ref{App_g11_CO}). Using the expressions from (\ref{ABD}), we find that in the case of GR ($\beta_0=\gamma_0=\xi_0=\delta_0=1$) the coefficient ${\cal A}G^{}_{11}/G^{}_{00}$ in Eq.~(\ref{App_g11_Will}) is equal to
\begin{equation}\label{AG}
\frac{{\cal A}G^{}_{11}}{G^{}_{00}}=1-\frac{\Phi}{c^2}+O\left(\frac{\Phi^2}{c^4}\right),
\end{equation}
while the coefficient ${\cal D}$ in Eq.~(\ref{App_g11_CO}) has another value:
\begin{equation}\label{D}
{\cal D}=1-\frac{3}{2}\frac{\Phi}{c^2}+O\left(\frac{\Phi^2}{c^4}\right),
\end{equation}
i.e. ${\cal D}\neq{\cal A}G^{}_{11}/G^{}_{00}$. We will also obtain a similar negative result by comparing the expressions for the time metric component $g^{}_{00}({\bf r})$ in the formulas (\ref{App_ds_CO_GR}) and (\ref{nonlinear_PN_loc}).

Thus, we have shown that the expression (\ref{App_ds_CO_GR}) can never be consistent with the basic expression (\ref{nonlinear_PN_loc}) obtained by solving of the Einstein equations directly in the cosmological reference frame $\{t,{\bf r}\}_\text{Cosm}$ without additional artificial assumptions. Therefore, formula (\ref{App_ds_LO_GR}) with the post-Newtonian coefficients $\beta =\gamma=1$ (in the laboratory reference frame $\{t',{\bf r}'\}_\text{Lab}$) is also not justified and cannot be used as an argument against the cosmological gravimetry developed by us, in which $\beta\neq1$ and $\gamma\neq 1$ due to the influence of the cosmological gravitational background.

\section{Cosmologically insensitive model of gravitation}
Below we consider some features of the cosmologically insensitive metric model of gravitation. First of all, we note that from a formal viewpoint there is a more general formulation than in Eq.~(\ref{g_noCG}), which also satisfies the Postulates~1-3:
\begin{equation}\label{g_noCG2}
g_{00}({\bf r})=e^{2U({\bf r})/c^2},\quad g_{11}({\bf r})=e^{-2wU({\bf r})/c^2},
\end{equation}
where $w$ is an arbitrary real coefficient. However, taking into account the known empirical data for PPN parameters ($\gamma\approx 1$), only the model with $w=1$ can be correct, i.e. Eq.~(\ref{g_noCG}).

\subsection{Chronometric gravimetry}
According to general relativity \cite{Einstein_1915}, the passage of time in a gravitational field slows down.
Consider the well-known gravitation-chronometric experiments (for example, see Refs.~\cite{Chou_Sc_2010,Katori_2016,Katori_2020}) on measuring of the ratio
\begin{equation}\label{v1v2}
\frac{\Delta\omega}{\omega}=\frac{\omega({\bf r}'_1)-\omega({\bf r}'_2)}{\omega({\bf r}'_1)}=1-\frac{\omega({\bf r}'_2)}{\omega({\bf r}'_1)}\,,
\end{equation}
which describes the relative difference between the frequencies $\omega({\bf r}'_1)$ and $\omega({\bf r}'_2)$ for the same atomic transition at two different spacial points ${\bf r}'_1$ and ${\bf r}'_2$ with different gravitational potentials $\varphi({\bf r}'_1)$ and $\varphi({\bf r}'_2)$. In the framework of the metric theory of gravitation, the following relation holds:
\begin{equation}\label{ratio_v1v2}
\frac{\omega({\bf r}_2)}{\omega({\bf r}_1)}=\frac{\omega({\bf r}'_2)}{\omega({\bf r}'_1)}=\sqrt{\frac{{g}_{00}({\bf r}_2)}{{g}_{00}({\bf r}_1)}}=\sqrt{\frac{\tilde{g}_{00}({\bf r}'_2)}{\tilde{g}_{00}({\bf r}'_1)}}\,.
\end{equation}
Then, using Eq.~(\ref{g_noCG_Lab}) for a cosmologically insensitive model, we obtain the expression:
\begin{align}\label{v1v2_noCG}
& \frac{\Delta\omega}{\omega}=1-e^{-\Delta\varphi/c^2}=\frac{\Delta\varphi}{c^2}-\frac{(\Delta\varphi)^2}{2c^4}+\frac{(\Delta\varphi)^3}{6c^6}-...\,,\\
& \Delta\varphi=\varphi({\bf r}'_1)-\varphi({\bf r}'_2)\,,\nonumber
\end{align}
which depends only on $\Delta\varphi$. This is due to the exponential dependence of the metric component $\tilde{g}_{00}$ on the potential $\varphi({\bf r}')$ [see Eq.~(\ref{g_noCG_Lab})], because ${\Delta\omega}/{\omega}$ depends only on the ratio ${\tilde{g}_{00}({\bf r}'_2)}/{\tilde{g}_{00}({\bf r}'_1)}$ [see Eqs.~(\ref{v1v2}) and (\ref{ratio_v1v2})]. While any other law of $\tilde{g}_{00}$ on the potential $\varphi({\bf r}')$ will lead to the dependence ${\Delta\omega}/{\omega}$, which cannot be expressed only in terms of $\Delta\varphi$.

For comparison, if we use for $\tilde{g}_{00}({\bf r}')$ the expression (\ref{nonlinear_PN_lab_2}) obtained in the framework of the PPN formalism, then up to quadratic contributions of $\varphi({\bf r}')$  we have the following result:
\begin{equation}\label{v1v2_PN}
\frac{\Delta\omega}{\omega}\approx\frac{\Delta\varphi}{c^2}-\frac{(\Delta\varphi)^2}{2c^4}+(\beta-1)\frac{\varphi^2({\bf r}'_1)-\varphi^2({\bf r}'_2)}{c^4}\,.
\end{equation}
This formula shows the principal possibility of cosmological gravimetry (i.e., if $\beta\neq 1$) in space-based chronometric experiments using ultra-precision atomic clocks for spacecraft at low and high orbits around the Sun.

\subsection{Free fall acceleration in nonlinear regime. Schwarzschild radius.}

\begin{figure}[t]
\centerline{\scalebox{0.5}{\includegraphics{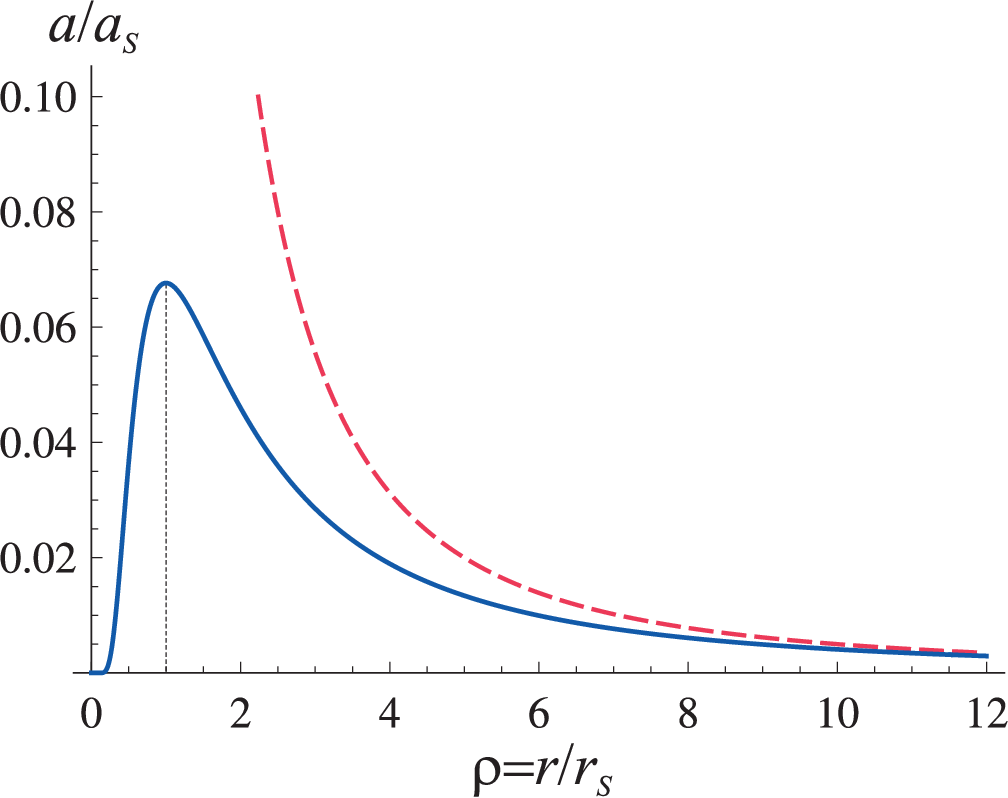}}}\caption{
Dependence of the free fall acceleration for the cosmologically insensitive model of gravitation (solid thick line) [see Eq.~(\ref{a2})]. The dashed line corresponds to the classical dependence for Newtonian gravity: $a=\Gamma M/r^2$.} \label{Fig2}
\end{figure}

Using expressions (\ref{g_noCG_Lab}), we write the equation for the geodesic line (\ref{geod_lab_gen}) in the nonrelativistic low-velocity limit:
\begin{equation}\label{geod_noCG}
{\bf a}=\frac{d^2 {\bf r}}{d t^2}=-e^{4\varphi({\bf r})/c^2}\nabla\varphi({\bf r})\,,
\end{equation}
which describes the free fall acceleration ${\bf a}$ for the cosmologically insensitive model of gravitation. For the point mass $M$ placed in the center of the spatial reference frame, the potential has the standard form: $\varphi({\bf r})=-\Gamma M/r$, where $r=|{\bf r}|$. We introduce the well-known Schwarzschild radius, $r^{}_S=2M\Gamma /c^2$, which allows us to express the gravitational potential as $\varphi({\bf r})=-c^2 r^{}_S/2r$. In this case, from (\ref{geod_noCG}) we obtain:
\begin{align}\label{a}
& {\bf a}=-e^{-2r_S/r}\frac{c^2 r^{}_S}{2r^2}{\bf n}\,,
\end{align}
where ${\bf n}={\bf r}/|{\bf r}|$ is the unit radial vector. We define the parameter $a^{}_S=c^2/r^{}_S$, which has the dimension of acceleration and which we will call as Schwarzschild acceleration. As a result, we obtain a universal expression for the value of $a=|{\bf a}|$ in the form of a dependence on the dimensionless parameter $\rho=r/r^{}_S$:
\begin{equation}\label{a2}
\frac{a}{a^{}_S}=\frac{e^{-2/\rho}}{2\rho^2}\,,
\end{equation}
which is presented in Fig.~\ref{Fig2}. This figure shows that there is an prominent maximum of $a_{max}=0.5\,e^{-2}a^{}_S\approx 0.068a^{}_S$ for $r=r^{}_S$, and at the distances $r<r^{}_S$ we see a sharp decrease in the gravitational effect. Thus, in spite of the fact that the cosmologically insensitive model of gravitation has no singularities at all (in contrast to GR for $r\leq r^{}_S$), nevertheless, the concept of Schwarzschild radius $r^{}_S$ for this model is also relevant, because the gravitational effect at the distances $r>r^{}_S$ is radically different from the behavior at the small distances, $r<r^{}_S$.

\subsection{Metric equivalent of Newtonian gravitation. Cosmological insensitivity.}
The metric equivalent of the classical Newtonian gravitation corresponds to the following expression for the interval:
\begin{equation}\label{int_New_1}
ds^2=\left(1+\frac{2U({\bf r})}{c^2}\right)(c\,dt)^2-(d{\bf r})^2\,.
\end{equation}
Indeed, in this case, the exact equation for the geodesic line (\ref{geod_x0}) has the form
\begin{equation}\label{Newt_eq}
{\bf a}=\frac{d^2 {\bf r}}{d t^2}=-\nabla_{\bf r}U({\bf r})\,,
\end{equation}
which is the standard equation of free motion in Newtonian theory.

Let us investigate the sensitivity of the metric formulation (\ref{int_New_1}) to the transformation law (\ref{invar}) which leads to the following expression
\begin{equation}\label{int_New_2}
ds^2=\left(1+\frac{2C_0}{c^2}+\frac{2U({\bf r})}{c^2}\right)(c\,dt)^2-(d{\bf r})^2\,.
\end{equation}
In this case, the transition to the laboratory space-time $\{t',{\bf r}'\}_\text{Lab}$ is carried out as follows:
\begin{equation}\label{New_LO}
dt'=dt\sqrt{1+2C_0/c^2}\,,\quad {\bf r}'={\bf r}\,,
\end{equation}
what results in
\begin{equation}\label{int_New_3}
ds^2=\left(1+\frac{2U({\bf r}')}{c^2(1+2C_0/c^2)}\right)(c\,dt')^2-(d{\bf r}')^2\,.
\end{equation}
In this case, the equation for the geodesic line (\ref{geod_x0}) has the form:
\begin{equation}\label{Newt_eq_2}
\frac{d^2 {\bf r}'}{d t'^2}=-\frac{1}{1+2C_0/c^2}\nabla_{{\bf r}'}U({\bf r}')\,.
\end{equation}
Comparing Eq.~(\ref{Newt_eq_2}) with the equation of free motion (\ref{mov_eq}), we find the expression for the gravitational potential $\varphi({\bf r}')$ from the viewpoint of the Laboratory Observer:
\begin{equation}\label{New_phi}
\varphi({\bf r}')=\frac{U({\bf r}')}{1+2C_0/c^2}\,.
\end{equation}
As a result, from Eq.~(\ref{int_New_3}) we get the final expression
\begin{equation}\label{int_New_4}
ds^2=\left(1+\frac{2\varphi({\bf r}')}{c^2}\right)(c\,dt')^2-(d{\bf r}')^2\,,
\end{equation}
which demonstrates the invariance of the metric expression (\ref{int_New_1}) with respect to the transformation law (\ref{invar}). However, this invariance is incomplete, because we have the restriction on the constant $C_0$:
\begin{equation}\label{C0}
C_0>-c^2/2\,,
\end{equation}
which follows from Eq.~(\ref{New_LO}). Indeed, if $C_0<-c^2/2$, then the coefficient $\sqrt{1+2C_0/c^2}$ in Eq.~(\ref{New_LO}) becomes purely imaginary, which has not physical meaning. While for the nonlinear PPN model (\ref{g_noCG}) there are no formal restrictions on the real value $C_0$ at all.

Thus, we believe that only the nonlinear PPN model (\ref{g_noCG}) and the linear Newtonian model (\ref{int_New_1}) are invariant to the transformation law (\ref{invar}), and therefore they are cosmologically insensitive. While any other metric models (including Einstein's general relativity) should contain the possibility of a cosmological gravimetry (at least, in principle).


\begin{thebibliography}{30}
\bibitem{we}
Note that the similar situation is not new in modern physics. For example, in the classical physics, the value of energy $E$ itself (for some considered system) is not really physical, because this theory is invariant to the transformation $E\rightarrow E+E_0$, where $E_0$ is an arbitrary real constant (in energy units). However, special relativity forbids this arbitrariness by the formula $E=Mc^2$, i.e., the energy $E$ itself becomes really physical value.

Therefore, why cannot it be admitted that a more advanced and nonlinear theory of gravitation (GR in our case) can forbid the transformation (\ref{invar}) allowed in the classical linear Newton's theory of gravity? This will lead to the fact that the value of the gravitational potential itself will also become really physical.
\end{thebibliography}
\end{document}